# Milagrito Detection of TeV Emission from Mrk 501

**Stefan Westerhoff** [1] **for the Milagro Collaboration**
[1] *University of California, Santa Cruz, CA 95064, USA*

**Abstract**

The Milagro water Cherenkov detector near Los Alamos, New Mexico, has been operated as a sky monitor at energies of a few TeV between February 1997 and April 1998. Serving as a test run for the full Milagro detector, Milagrito has taken data during the strong and long-lasting 1997 flare of Mrk 501. We present results from the analysis of Mrk 501 and compare excess and background rate with expectations from the detector simulation.

## 1 Introduction:

With the detection of 4 Galactic and 3 extragalatic sources, Very High Energy (VHE) $\gamma$-ray astronomy, studying the sky at energies above 100 GeV, has become one of the most interesting frontiers in astronomy. Source detections and analyses in this field are still dominated by the highly successful atmospheric Cherenkov technique. Cherenkov telescopes and telescope arrays are optimal tools for the detailed study of established sources and their energy spectra and the theory-guided search for yet unknown sources. There is, however, also a strong case for instruments able to perform an unbiased, systematic and continuous search for TeV sources, thus overcoming the limitations imposed by the low duty cycle and small field of view of Cherenkov telecopes. Consequently, the observation technique must exploit the *particle content* of air showers rather than the Cherenkov light.

A first-generation all-sky monitor operating at energies below 1 TeV, the Milagro detector (McCullough et al., 1999) located 2650 m above sea level near Los Alamos, New Mexico, at latitude $\lambda = 35.9°$ N, started data taking in early 1999. Milagro is a water Cherenkov detector of size $60 \times 80 \times 8 \, \text{m}^3$. Two layers of photomultiplier tubes detect the Cherenkov light produced by secondary particles entering the water. The first layer, with 450 tubes on a $3 \times 3 \, \text{m}^2$ grid at a depth of 1.4 m, allows the shower direction and thus the direction of the primary particle to be reconstructed, while the second layer with 273 tubes at a depth of $\sim$7.0 m primarily detects the penetrating component of air showers, *i.e.* muons, hadrons, and highly energetic electromagnetic particles.

A smaller, less sensitive prototype, Milagrito (Atkins et al., 1999), has taken data between February 1997 and April 1998. Milagrito, a one-layer detector of size $35 \times 55 \times 2 \, \text{m}^3$ with 228 photomultiplier tubes on a $3 \times 3 \, \text{m}^2$ grid at a rather shallow depth of 0.9 m, served mainly as a test run for this relatively new detection technique. This prototype has, however, taken data during a very intense and long-lasting flare of Mrk 501 in 1997 (Samuelson et al., 1998).

For the evaluation of the performance of VHE instruments, the Crab nebula is usually used as a "standard candle". It is a well-studied steady source with a flux of

$$J_\gamma(E) = (3.20 \pm 0.17 \pm 0.6) \times 10^{-7} E_{\text{TeV}}^{-2.49 \pm 0.06 \pm 0.04} \, \text{m}^{-2} \, \text{s}^{-1} \, \text{TeV}^{-1}, \qquad (1)$$

(Hillas et al., 1998). Simulations indicate that the expected significance from Milagrito for the Crab nebula is less than $2\,\sigma$, ruling out the possibility of using a Crab signal to test Milagrito's performance. A detection of Mrk 501 with a sufficiently high significance can be expected had the average flux been in excess of the Crab flux. During its flare in 1997, Mrk 501 has been intensively studied with several air Cherenkov telescopes. Although not covering the same observation times, the average fluxes measured by Whipple (Samuelson et al., 1998) and the HEGRA stereo system of air Cherenkov telescopes (Aharonian et al., 1999) agree extremely well both in shape and magnitude, and they both indicate a significant deviation of the energy spectrum from a simple power law. Using an average flux as measured by Whipple,

$$J_\gamma(E) = (8.6 \pm 0.3 \pm 0.7) \times 10^{-7} E_{\text{TeV}}^{-2.20 \pm 0.04 \pm 0.05 - (0.45 \pm 0.07)\log_{10} E} \, \text{m}^{-2} \, \text{s}^{-1} \, \text{TeV}^{-1}, \qquad (2)$$

simulations of the Milagrito detector response predict the expected integral $\gamma$-rate from Mrk 501 to be 3.6 times the Crab rate. Although highly variable sources like Mrk 501 are not well-suited for checking the sensitivity of detectors integrating over long time periods, the observation of an excess from Mrk 501 still provides a test for the sensitivity of Milagrito and reliability of the detector simulation.

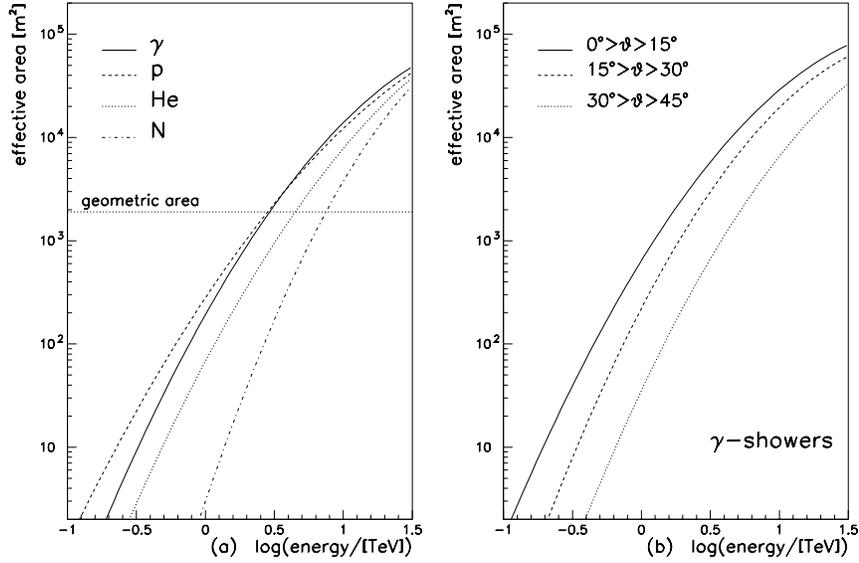

Figure 1: (a) Effective area of Milagrito for reconstructed $\gamma$- and cosmic ray showers, averaged over a zenith angle range from $0° \leq \theta \leq 45°$, as a function of the primary energy. (b) Effective area for $\gamma$-showers for various zenith angle ranges.

In addition, observations with Cherenkov telescopes cover only the time from February to October, while Milagrito continued to monitor Mrk 501 in late 1997 and early 1998.

## 2 Milagrito Performance:

Sensitivity predictions for Milagrito are based on a detector simulation using the CORSIKA 5.61 air shower simulation code for the development of the shower in the Earth's atmosphere, and the GEANT 3.21 package for the simulation of the detector. The simulation is described in detail elsewhere (Atkins et al.,1999).

The Milagrito detector operated with a minimum requirement of 100 hit tubes per event. Figure 1 (a) shows the effective area $A_{\text{eff}}$ of Milagrito for $\gamma$-showers and cosmic ray background showers induced by protons, helium, and nitrogen, the latter used for representing the combined CNO flux, as a function of the energy of the primary particle. Figure 1 (b) shows how the efficiency depends on the zenith angle $\theta$.

At energies $\leq 2\,\text{TeV}$, the effective area for proton-induced showers is larger than for $\gamma$-showers. This is related to the fact that $\gamma$-induced (thus almost purely electromagnetic) showers are usually more laterally confined so the area covered by the particles reaching detector altitude is smaller than for hadron-induced showers, which tend to have "hot spots" with high particle density at large distances from the shower core. At energies above $\sim 5\,\text{TeV}$, the larger effective area for $\gamma$-induced showers provides an intrinsic cosmic ray background rejection.

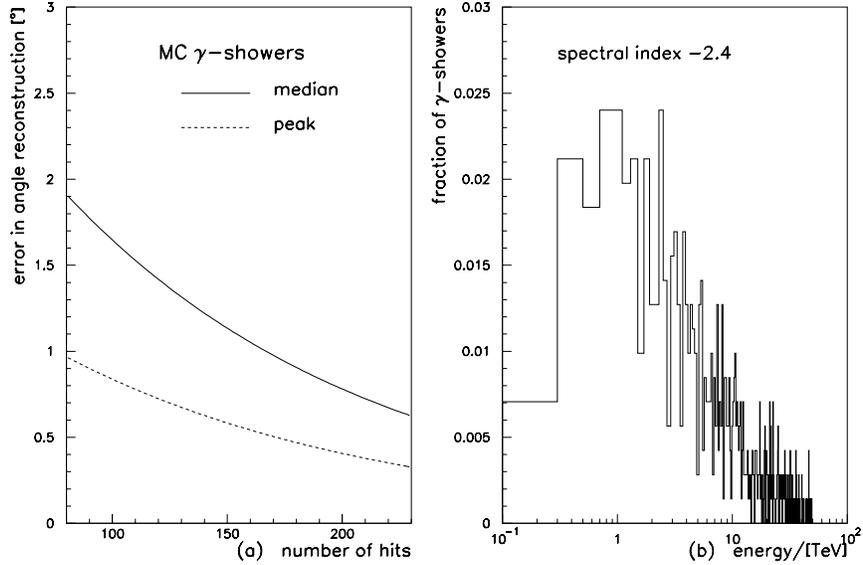

Figure 2: (a) Angular resolution and (b) energy distribution of $\gamma$-showers triggering Milagrito.

As a large fraction of showers fulfilling the trigger condition have their core outside the sensitive detector area, the effective area is larger than the geometrical area above $\sim 3\,\text{TeV}$. In fact, only $16\,\%$ of the proton showers and $21\,\%$ of the $\gamma$-showers triggering the Milagrito detector have their core on the pond. This leads to a rather broad energy distribution starting at energies as low as $100\,\text{GeV}$, with no well defined threshold energy (Figure 2 (b)). The median energy varies slightly with the source declination $\delta$, ranging from $\sim 3\,\text{TeV}$ for sources at $\delta = 40°$ to $7\,\text{TeV}$ for sources with $|\delta - \lambda| \simeq 20°$.

The water Cherenkov technique uses water both as the converter and the detector medium. Consequently, the efficiency for detecting low energy air shower particles is very high, leading to a good sensitivity even for showers with primary energy below $1\,\text{TeV}$. The angle fitting, however, has to deal with a considerable amount of light late as compared to the shower front reaching the detector. The "late light" is partly produced by low energy particles which tend to trail the shower front. More important, however, is the horizontal light component resulting from the large Cherenkov angle in water ($41°$), multiple scattering, $\delta$-rays, and scattering and reflection of Cherenkov light. The expected angular resolution for cosmic rays agrees with our observations of the cosmic ray shadow of the moon (Wascko et al., 1999).

Milagrito's angular resolution is a strong function of the number of the tubes in the fit to the arrival times of the tubes (Figure 2 (a)). For the initial source search, a minimum number of 40 tubes used in the shower plane fit is required. This leads to a measured rate of $2950 \pm 98$ reconstructed events per day from cosmic ray showers in a typical source bin with $1.1°$ radius at the declination of Mrk 501. This is in good agreement with the predicted rate of $3080^{+205}_{-110}$ events per day from protons, Helium, and CNO nuclei. In the simulation, the contribution of He and CNO to the total trigger rate turns out to be $27\,\%$ and $4\,\%$, respectively.

## 3 Results:

A straight-forward analysis with a source bin of radius $1.1°$ centered on Mrk 501 leads to an excess $> 3\,\sigma$. According to simulations this bin size contains $48\,\%$ of the source events and is optimal for an analysis treating all events equally. The corresponding excess rate averaged over the lifetime of Milagrito (370 equivalent source days for Mrk 501) is $(8.7 \pm 3.0)\,\text{day}^{-1}$.

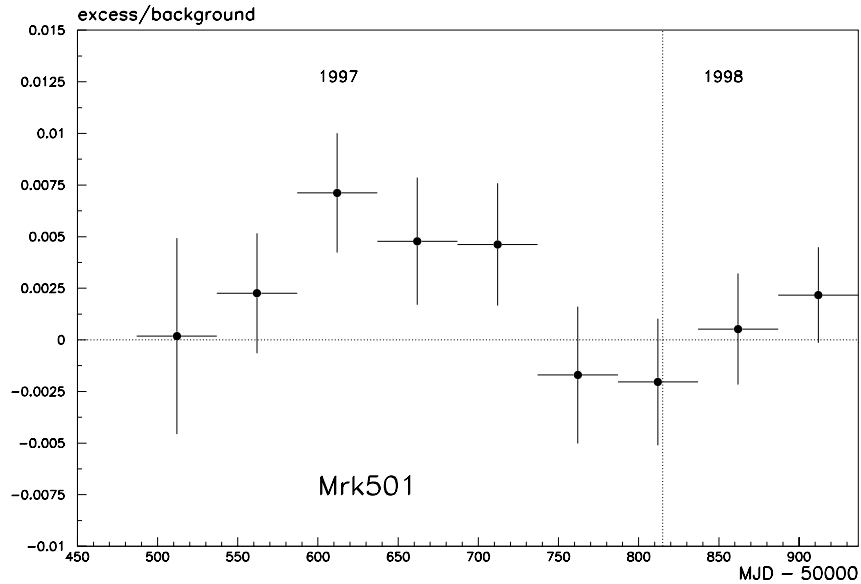

Figure 3: Excess/background for Mrk 501 as a function of time. At the current sensitivity level the data is consistent with a constant flux.

Figure 3 shows how the excess is accumulated over Milagrito's lifetime. At our present level of sensitivity, the data is consistent with a flux constant in time.

Using the average flux as measured by air Cherenkov telescopes between February and October 1997, simulations predict a $\gamma$-rate of $(13.3 \pm 4.0)\,\mathrm{day}^{-1}$ for Milagrito and are thus consistent with the measured excess during this period, $(15.3 \pm 4.6)\,\mathrm{day}^{-1}$.

An analysis that takes account of the strong dependence of the resolution on the number of photomultipliers in the fit should be more sensitive to emission from a point source. The results of such an analysis will be presented at the conference.

The analysis was extended to 10 other nearby blazars ($z \leq 0.06$) in Milagrito's field of view, but Mrk 501 remains the only analyzed source with a significance in excess of $3\,\sigma$. Results from this blazar sample are reported elsewhere (Westerhoff et al., 1999).

This research was supported in part by the National Science Foundation, the U. S. Department of Energy Office of High Energy Physics, the U. S. Department of Energy Office of Nuclear Physics, Los Alamos National Laboratory, the University of California, the Institute of Geophysics and Planetary Physics, The Research Corporation, and the California Space Institute.